\documentclass{article}

\usepackage[utf8]{inputenc}
\usepackage{enumitem}
\usepackage{listings}
\usepackage[T1]{fontenc}
\usepackage{multirow}
\usepackage[utf8]{inputenc}

\usepackage{spconf,amsmath,graphicx}

\usepackage{microtype}
\usepackage{graphicx}

\usepackage{subcaption}

\usepackage{booktabs} 
\usepackage{xcolor} 
\usepackage{pgfplots,pgfplotstable}

\usepackage{hyperref}
\usepackage{url}

\usepackage{amsmath}
\usepackage{amssymb}
\usepackage{mathtools}
\usepackage{amsthm}
\usepackage{bm}
\usepackage[capitalize,noabbrev]{cleveref}

\definecolor{forestgreen}{rgb}{0.13, 0.55, 0.13}
\definecolor{fulvous}{rgb}{0.86, 0.52, 0.0}
\definecolor{glaucous}{rgb}{0.38, 0.51, 0.71}
\definecolor{lava}{rgb}{0.81, 0.06, 0.13}
\definecolor{buff}{rgb}{0.94, 0.86, 0.51}
\definecolor{chromeyellow}{rgb}{1.0, 0.65, 0.0}
\definecolor{brightube}{rgb}{0.82, 0.62, 0.91}

\hypersetup{
    pdfborderstyle={/S/U/W 1}, 
    linkbordercolor=red,       
    citebordercolor=green,     
    filebordercolor=magenta,   
    urlbordercolor=cyan        
}

\pgfplotsset{scaled y ticks=false, scaled x ticks=false}

\theoremstyle{plain}

\theoremstyle{definition}

\theoremstyle{remark}

\usepackage{xspace}

\usepackage[textsize=tiny]{todonotes}

\usepackage{siunitx}

\usepackage{float}
\usepackage{adjustbox}

\title{VANI: Very-lightweight Accent-controllable TTS for Native and Non-native speakers with Identity Preservation}

\name{
    \begin{tabular}{c c c c }
    Rohan Badlani & Akshit Arora & Subhankar Ghosh & Rafael Valle \\
    Kevin J. Shih & João Felipe Santos & Boris Ginsburg & Bryan Catanzaro
    \end{tabular}
}

\address{NVIDIA}
\begin{document}

\maketitle

\section{Abstract}
\vspace*{-0.75\baselineskip}
We introduce VANI, a very lightweight multi-lingual accent controllable speech synthesis system. Our model builds upon disentanglement strategies proposed in RADMMM\cite{badlani2023multilingual} and supports explicit control of accent, language, speaker and fine-grained $F_0$ and energy features for speech synthesis. We utilize the Indic languages dataset, released for LIMMITS 2023 as part of ICASSP Signal Processing Grand Challenge, to synthesize speech in 3 different languages. Our model supports transferring the language of a speaker while retaining their voice and the native accent of the target language. We utilize the large-parameter RADMMM model for Track $1$ and lightweight VANI model for Track $2$ and $3$ of the competition. 

\vspace*{-1.0\baselineskip}
\section{Introduction}
\vspace*{-0.75\baselineskip}
There has been incredible progress in the quality of text-to-speech(TTS) models. However, most TTS models do not disentangle attributes of interest. Our goal is to create a multi-lingual TTS system that can synthesize speech in any target language (with the target language's native accent) for any speaker seen by the model. The main challenge is disentanglement of attributes like speaker, accent and language such that the model can synthesize speech for any desired combination of these attributes without any bi-lingual data. 

\vspace*{-1.0\baselineskip}
\section{Method}
\vspace*{-0.75\baselineskip}
\subsection{Dataset and Preprocessing}
\vspace*{-0.7\baselineskip}
\label{sec:dataset}
We utilize the Hindi, Telugu, and Marathi dataset released as part of LIMMITS challenge. We remove empty audio files and clips with duplicate transcripts. We parse files through Automatic Speech Recognition model and generate transcripts. We select top $8000$ datapoints per speaker with the least character error rate (CER) between ground truth and generated transcripts. This results in the dataset used for Track $2$. For Track $1$ and $3$, we identify audio clips with maximal overlap in characters across speakers within a language\footnote{Dataset and Model Parameter Details: https://bit.ly/icassp\_vani}. We trim the leading and trailing silences and normalize audio volume.

\begin{table*}[h]
\caption{\label{tab:vanimetrics}Mono-lingual Evaluation using Cosine Sim(speaker identity retention) and CER(content quality) on resynthesis.}
\vspace*{-0.5\baselineskip}
    \centering
    \begin{adjustbox}{max width=\textwidth}
    \begin{tabular}{|l|l|l|l|l|l|l|l|l|l|l|l|l|}
    \hline
    Model & \multicolumn{2}{|c|}{Hindi Female} & \multicolumn{2}{|c|}{Hindi Male} &  \multicolumn{2}{|c|}{Marathi Female} & \multicolumn{2}{|c|}{Marathi Male} & \multicolumn{2}{|c|}{Telugu Female} & \multicolumn{2}{|c|}{Telugu Male} \\ \hline
         ~ & Cosine Sim ($\uparrow$) & CER ($\downarrow$) & Cosine Sim ($\uparrow$) & CER ($\downarrow$) & Cosine Sim ($\uparrow$) & CER ($\downarrow$) & Cosine Sim ($\uparrow$) & CER ($\downarrow$) & Cosine Sim ($\uparrow$) & CER ($\downarrow$) & Cosine Sim ($\uparrow$) & CER ($\downarrow$)\\ \hline
          GT & 1.0 & 0.035 & 1.0 & 0.015 & 1.0 & 0.089 & 1.0 & 0.094 & 1.0 & 0.049 & 1.0 & 0.049\\  \hline
        Track1 (RADMMM + Aug + HiFiGAN) & $ 0.865 \pm 0.022 $ & 0.045 & $ 0.871 \pm 0.022 $ & 0.015 & $ 0.809 \pm 0.045 $ & 0.096 & $ 0.869 \pm 0.016 $ & 0.127 & $ 0.831 \pm 0.056 $ & 0.056 & $ 0.856 \pm 0.027 $ & 0.047\\ \hline  
        Track2 (VANI-NP + Waveglow) & $ 0.853 \pm 0.025 $ & 0.072 & $ 0.773 \pm 0.037 $ & 0.034 & $ 0.757 \pm 0.052 $ & 0.221 & $ 0.785 \pm 0.031 $ & 0.243 & $ 0.770 \pm 0.071 $ & 0.124 & $ 0.758 \pm 0.026 $ & 0.111 \\ \hline 
         Track2 (VANI-NP + HiFiGAN) & $ 0.829 \pm 0.027 $ & 0.079 & $ 0.740 \pm 0.028 $ & 0.036 & $ 0.727 \pm 0.052 $ & 0.211 & $ 0.727 \pm 0.036 $ & 0.242 & $ 0.750 \pm 0.041 $ & 0.112 & $ 0.727 \pm 0.076 $ & 0.148 \\ \hline 
        Track3 (VANI-P + Aug + Waveglow) & $ 0.842 \pm 0.037 $ & 0.094 & $ 0.782 \pm 0.013 $ & 0.043 & $ 0.740 \pm 0.044 $ & 0.224 & $ 0.760 \pm 0.034 $ & 0.256 & $ 0.767 \pm 0.053 $ & 0.156 & $ 0.706 \pm 0.037 $ & 0.172 \\ \hline 
         Track3 (VANI-P + Aug + HiFiGAN) & $ 0.845 \pm 0.018 $ & 0.067 & $ 0.758 \pm 0.052 $ & 0.045 & $ 0.745 \pm 0.031 $ & 0.229 & $ 0.759 \pm 0.042 $ & 0.233 & $ 0.771 \pm 0.032 $ & 0.151 & $ 0.701 \pm 0.042 $ & 0.200 \\
        
    \bottomrule
    \end{tabular}
    \end{adjustbox}
\end{table*}

\begin{table*}[h]
\caption{\label{tab:vanitransfermetrics}Cross-Lingual Evaluation on 2 languages using Cosine Sim(speaker identity retention) and CER(content quality).}
\vspace*{-0.5\baselineskip}
    \centering
    \begin{adjustbox}{max width=\textwidth}
    \begin{tabular}{|l|l|l|l|l|l|l|l|l|l|l|l|l|}
    \hline
    Model & \multicolumn{2}{|c|}{Hindi Female} & \multicolumn{2}{|c|}{Hindi Male} &  \multicolumn{2}{|c|}{Marathi Female} & \multicolumn{2}{|c|}{Marathi Male} & \multicolumn{2}{|c|}{Telugu Female} & \multicolumn{2}{|c|}{Telugu Male} \\ \hline

         ~ & Cosine Sim ($\uparrow$) & CER ($\downarrow$) & Cosine Sim ($\uparrow$) & CER ($\downarrow$) & Cosine Sim ($\uparrow$) & CER ($\downarrow$) & Cosine Sim ($\uparrow$) & CER ($\downarrow$) & Cosine Sim ($\uparrow$) & CER ($\downarrow$) & Cosine Sim ($\uparrow$) & CER ($\downarrow$)\\ \hline
        Track1 (RADMMM + Aug + HiFiGAN) &  $ 0.295 \pm 0.060 $ & 0.041 & $0.324 \pm 0.100$ & 0.016 & $0.339 \pm 0.117 $ & 0.138 & $0.301 \pm 0.071 $ & 0.160 & $0.352 \pm 0.105$ & 0.06 & $ 0.301 \pm 0.076 $ & 0.048\\ \hline 
         Track2 (VANI-NP + HiFiGAN) & $0.273 \pm 0.081 $ & 0.072 & $0.294 \pm 0.107 $ & 0.040 & $0.330 \pm 0.128 $ & 0.268 & $0.305 \pm 0.112 $ & 0.276 & $0.313 \pm 0.125$ & 0.136 &  
$0.299 \pm 0.115 $ & 0.142 \\ \hline
        Track3 (VANI-P + Aug + HiFiGAN) & $0.288 \pm 0.072 $ & 0.071 & $0.266 \pm 0.104 $ & 0.045 & $ 0.308 \pm 0.142 $ & 0.255 & $0.307 \pm 0.150 $ & 0.265 & $0.285 \pm 0.139 $ & 0.203 & $0.282 \pm 0.142$ & 0.218 \\ 
        
    \bottomrule
    \end{tabular}
    \end{adjustbox}
\end{table*}

\begin{table}[h]
\caption{\label{tab:leaderboard} LIMMITS'23 Competition human evaluation results.}
\vspace*{-0.5\baselineskip}
    \centering
    \resizebox{\columnwidth}{!}{\begin{tabular}{|l|l|l|}
    \hline
        ~ & Naturalness ($\uparrow$) & Speaker Similarity ($\uparrow$) \\ \hline
        Track 1 (RADMMM + Aug + HiFiGAN) & $4.71$ & $3.98$ \\ \hline
        Track 2 (VANI-NP + Waveglow) & $4.12$ & $3.02$ \\ \hline
        Track 3 (VANI-P + Aug + Waveglow) & $4.04$ & $2.76$ \\ \hline
    \end{tabular}}
\end{table}

\vspace*{-0.5\baselineskip}
\subsection{Spectogram Synthesis Model}
\vspace*{-0.5\baselineskip}
Our goal is to develop a model for multilingual synthesis in the languages of interest with the ability of cross-lingual synthesis for a speaker of interest. Our dataset comprises of each speaker speaking only one language and hence there are correlations between text, language, accent and speaker within the dataset. Recent work on RADMMM~\cite{badlani2023multilingual} tackles this problem by proposing several disentanglement approaches. We utilize RADMMM as the base model for track $1$. For track $2, 3$ we use the proposed lightweight VANI model. As in RADMMM, we use deterministic attribute predictors to predict fine-grained features given text, accent and speaker.

We leverage the text pre-processing, shared alphabet set and the accent-conditioned alignment learning mechanism proposed in RADMMM to our setup. This supports code-switching by default. We consider language to be \emph{implicit in the phoneme sequence} whereas the information captured by accent should explain the fine-grained differences between \emph{how phonemes are pronounced in different languages}.

\vspace*{-0.75\baselineskip}
\subsection{Track1: Large-parameter setup, Small-data}
\vspace*{-0.5\baselineskip}
As described in Sec \ref{sec:dataset}, our dataset is limited to $5$ hours per speakers. Since our dataset is very limited, we apply formant scaling augmentation suggested in RADMMM\cite{badlani2023multilingual} with the goal of disentangling speaker $S$ and accent $A$ attributes. We apply constant formant scales of $0.875$ and $1.1$ to each speech sample to obtain 2 augmented samples and treat those samples belonging to 2 new speakers. This helps reduce correlation between speaker, text and accent by having the those variables same for multiple speakers and provides more training data. Our model synthesizes mels($X \in \mathbb{R}^{C_{\mathit{mel}} \times F}$) using encoded text($\Phi \in \mathbb{R}^{C_{\mathit{txt}} \times T}$), accent($A \in \mathbb{R}^{D_{\mathit{accent}}}$), speaker($S \in \mathbb{R}^{D_{\mathit{speaker}}}$), fundamental frequency($F_0 \in \mathbb{R}^{\mathit{1} \times F}$) and energy($\xi \in \mathbb{R}^{\mathit{1} \times F}$) as conditioning variables where $F$ is the number of mel frames, $T$ is the text length, and energy is per-frame mel energy average. Although we believe attribute predictors can be generative models, we use deterministic predictors where $F_0^h$, $\mathcal{E}^h$ and $\Lambda^h$ are predicted pitch, energy, and durations conditioned on text $\Phi$, accent $A$ and speaker $S$:  

\begin{equation}
    P_{vani}(X) = P_{mel}(X | \Phi, \Lambda^h, A, S, F_0^h, \mathcal{E}^h)
\end{equation}

\vspace*{-0.75\baselineskip}
\subsection{Track2: Small-parameter, Large-data setup}
\vspace*{-0.5\baselineskip}
Since our goal is to have very lightweight model ($<5$ million parameters), we replace RADMMM mel-decoder with an autoregressive architecture. Our architecture is very similar to Flowtron\cite{Flowtron} with 2 steps of flow (one forward and one backward). Each flow step uses 3 LSTM layers and is conditioned on text, accent, speaker, $F_0$ and $\xi$. 

\vspace*{-0.75\baselineskip}
\subsection{Track3: Small-parameter, Small-data setup}
\vspace*{-0.5\baselineskip}
We utilize the model from Track $2$ and the data and augmentation strategy from Track $1$ as the model and data for Track $3$.

\vspace*{-0.75\baselineskip}
\subsection{Vocoder}
\vspace*{-0.5\baselineskip}
We use the HiFiGAN\footnote{NeMo implementation of HiFi-GAN: github.com/NVIDIA/NeMo} for Track $1$ and Waveglow\footnote{Waveglow checkpoints: github.com/bloodraven66/ICASSP\_LIMMITS23} for Track 2 and 3 to convert mel-spectrograms to waveforms. 

\vspace*{-0.75\baselineskip}
\section{Results and Analysis}
\vspace*{-0.5\baselineskip}
In this section, we evaluate the performance of the models in terms of content quality and speaker identity retention. 

\vspace*{-1.0\baselineskip}
\subsection{Character Error Rate (CER):}
\vspace*{-0.75\baselineskip}
We calculate CER between the transcripts generated from synthesized speech and ground truth(GT) transcripts. Models with lower CER are better in terms of content quality.

\vspace*{-1.0\baselineskip}
\subsection{Speaker Embedding Cosine Similarity:}
\vspace*{-0.75\baselineskip}
We use Titanet\cite{titanet} to get speaker embeddings and compare cosine similarity of synthesized sample against GT samples for same speaker. Higher scores show better identity retention.

\vspace*{-1.0\baselineskip}
\subsection{Evaluation Task Definition}
\vspace*{-0.75\baselineskip}
Table \ref{tab:vanimetrics} compares the Track1 model (RADMMM) against Track2 (VANI with nonparallel dataset - VANI-NP) and Track3 (VANI with limited parallel dataset - VANI-P) on mono-lingual resynthesis of speakers on 10 prompts in their own language (resynthesis task). Table \ref{tab:vanitransfermetrics} compares the models in the three tracks where speech was synthesized in a speaker's voice on 50 prompts outside of their own language (transfer task). 

\vspace*{-0.75\baselineskip}
\subsection{Analysis}
\vspace*{-0.5\baselineskip}
We observe that even with the limited dataset, the large parameter RADMMM model outperforms small parameter VANI model. We notice that Track2 with a larger dataset retains identity and content quality better than Track3 with limited data. However, all tracks do reasonably well on maintaining identity. We observe that on transfer, we're able to achieve decent CER comparable to the resynthesis case, indicating our model preserves content on transferring language of the speaker. The identity retention in transfer task is worse than resynthesis as expected but doesn't degrade much in VANI as compared to RADMMM demonstrating the importance of disentanglement strategies. We observe similar trend across tracks with human evaluation metrics (Table \ref{tab:leaderboard}).

\vspace*{-0.5\baselineskip}
\vspace*{-0.5\baselineskip}
\section{Conclusion}
\vspace*{-0.5\baselineskip}
We utilize strategies proposed in RADMMM~\cite{badlani2023multilingual} to disentangle speaker, accent and text for high-quality multilingual speech synthesis. We also present VANI, a lightweight multilingual autoregressive TTS model. We utilize several data preprocessing and augmentation strategies to preserve speaker identity in cross-lingual speech synthesis. Our model(s) can synthesize speech with proper native accent of any target language for any seen speaker without relying on bi-lingual data.

\bibliographystyle{IEEEbib}
\bibliography{main}\label{sec:references}

\end{document}